\documentstyle[pre,aps,preprint,epsf]{revtex} 

\newcommand{\abc}
  {\hbox{$\mathrm{A} + \mathrm{B} \rightleftharpoons \mathrm{C}$}}

\newcommand{\app}{\rightarrow}

\newcommand{\cald}{D}
\newcommand{\cale}{E}
\newcommand{\calf}{F}
\newcommand{\dee}{\partial}

\newcommand{\delf}{\Delta F}

\newcommand{\eps}{\varepsilon}
\newcommand{\exrm}{\mathrm{Ex}}

\newcommand{\eq}{{\text{eq}}}
\newcommand{\fbar}{\bar{f}}

\newcommand{\kinf}{K_\infty}
\newcommand{\kt}{K(T)}
\newcommand{\ktwid}{\tilde{K}}

\newcommand{\lbrak}{\left [}
\newcommand{\lpar}{\left (}

\newcommand{\ph}{}
\newcommand{\rbrak}{\right ]}

\newcommand{\rha}{\rho_{\rma}}
\newcommand{\rhb}{\rho_{\rmb}}
\newcommand{\rhc}{\rho_{\rmc}}
\newcommand{\rhceq}{\rho_c^{\text{eq}}}
\newcommand{\rhomax}{\rho_{\mathrm max}}
\newcommand{\rhostar}{\rho^*}

\newcommand{\rlharp}{\rightleftharpoons}
\newcommand{\rma}{a}
\newcommand{\rmb}{b}
\newcommand{\rmc}{c}

\newcommand{\rpar}{\right )}

\newcommand{\tstar}{T^*}
\newcommand{\uu}{u}
\newcommand{\vv}{v}
\newcommand{\ww}{w}
\newcommand{\weq}{w_{\text{eq}}}

\newcommand{\xxy}
  {\hbox{$2 {\mathrm{A}} \rightleftharpoons {\mathrm{C}}$}}

\newcommand{\boxhalf}{\hbox{$\frac{1}{2}$}}

\newcommand{\fftz}{F_{20}}
\newcommand{\ffoo}{F_{11}}
\newcommand{\ffzt}{F_{02}}

\newcommand{\eeto}{E_{21}}
\newcommand{\eeot}{E_{12}}

\newcommand{\ddozo}{D_{101}}
\newcommand{\ddzoo}{D_{011}}

\newcommand{\cplsec}
  {\vspace{.5in} \stepcounter{sec} \noindent \textbf{\arabic{sec}. } \textbf}

\begin{document}

\newcounter{sec}
\newcounter{subsec}
\newcounter{app}
\setcounter{sec}{0}
\setcounter{subsec}{0}
\setcounter{app}{0}

\title{Chemical Association via Exact Thermodynamic Formulations} 
\author{Michael E. Fisher and Daniel M. Zuckerman}
\address{Institute for Physical Science and Technology,\\ 
University of Maryland, College Park, Maryland 20742}
\date{\today}
\maketitle

\begin{abstract}
\hspace{.2in} 
It can be fruitful to view two-component physical systems of attractive monomers, A and B, ``chemically'' in terms of a reaction \abc $\mbox{}$, where C = AB is an associated pair or complex.
We show how to construct free energies in the three-component or chemical picture which, under mass-action equilibration, {\em exactly} reproduce {\em any} given two-component or ``physical'' thermodynamics.
Order-by-order matching conditions and 
{\em closed-form} chemical representations reveal the freedom available to modify the A-C, B-C, and C-C interactions and to adjust the association constant.
The theory (in the simpler one-component, i.e., A $\equiv$ B, case) is illustrated by treating a van der Waals fluid.
\end{abstract}
\pagebreak

\cplsec{Introduction and background}

The clustering of atoms and molecules in equilibrium is a nearly ubiquitous experimental phenomenon, occurring even among neutral, symmetric atoms like argon \cite{argon1,argon2}.
Not surprisingly, then, theoretical investigation of chemical association has remained a basic topic in physical chemistry since the work of Dolezalek \cite{dolezalek} early this century.
Advances in microscopically-based statistical mechanical descriptions include those of Hill \cite{hill} and Wertheim \cite{wertheim}.
Of greater interest here, however, are the more practical ``thermodynamic'' approximations, such as those of 
Heidemann and Prausnitz \cite{heidpraus},
who proposed a closed-form analytical approximation embodying all cluster sizes, and of Ebeling \cite{falkeb,ebgr} whose approach for electrolytes matched exactly known virial coefficients.
More recent treatments are discussed by Anderko \cite{anderko} and Economou and Donohue \cite{economou}.

The present discussion is limited to chemical formulations describing only ``pairing'' and employing a single association constant $K(T) \equiv K_2(T)$.
In the context of electrolyte theory --- which serves to illustrate {\em general} characteristics of chemical approximations --- 
the importance of dipolar pairing was first stressed by Bjerrum \cite{bjerrum}, and has proved crucial \cite{dmzfl} in describing the critical region of the ``restricted primitive model'' \cite{fl,lf}.
In electrolytes, as in other systems, one expects cluster-inducing attractions to dominate at low temperatures, $T$:
loosely speaking, clustering descriptions attempt to represent the partial ``saturation'' of such interactions, insofar as clustered particles are supposed to be in close contact.

The theoretical description of the {\em temperature variation} of association presents serious challenges and entails basic restrictions on the form of the association constant \cite{dmzfl,fl,lf,fzlong}.
As $T$ rises and excluded-volume effects dominate the thermodynamics, one expects heuristically the degree of clustering to drop and eventually to vanish at high $T$.
However, with Lee \cite{dmzfl} we have recently shown that electrolyte pairing theories with association constants, $\kt$, which vanish at high $T$ yield {\em negative} specific heats and so violate thermodynamic convexity requirements  (resting on the Second Law of Thermodynamics \cite{callen}) for significant regimes of temperature and density, $\rho$.
Moreover, the underlying mechanism \cite{dmzfl,fzlong} proves {\em not} specific to electrolyte theories.
On the other hand, acceptable forms, like the simple expression
$\kt \equiv K_\infty e^{T_0/T}$, do not vanish at high $T$; 
but the resulting finite value, $K=K_\infty$ when $T \app \infty$, then imposes {\em some} residual pairing even when repulsive forces totally dominate.
Chemical descriptions employing appropriate expressions for $\kt$ are, therefore, forced to accommodate pairing even for hard-core particles at infinite temperature!

A pairing  description of hard-core particles, in addition to seeming rather unphysical, is not easy to accomplish {\em accurately}.
Consider a model of attractive, spherical monomers, A and B, with hard-core repulsions.
A fairly good ``physical'' equation of state for the monomers when $T \app \infty$ is provided by the Carnahan-Starling (CS) theory and its extensions \cite{cs,mansoori}.
However, once C = AB pairs are also considered to be present in mass-action equilibrium at high $T$, as implied by $\kinf = K(T \app \infty) > 0$, no comparably accurate {\em chemical} hard-core descriptions seem available.
(See \cite{fzlong} for further details.)

These considerations lead to the basic question addressed here:
``How can one incorporate into an associative or chemical framework as much of the exact thermodynamics as is reliably known, in a tractable way that does not violate thermodynamic convexity requirements?''
In particular, one wants to maintain the accuracy of the monomeric repulsive-core physics without violating thermodynamic principles; 
but also desires to incorporate known virial or high-$T$ expansions.
In all cases, one hopes to benefit at low temperatures from the associative, chemical description which, at least at lower densities, becomes increasingly realistic.  
(See, e.g., \cite{lf}.)

In pursuit of these goals, we have studied systems consisting of two distinct monomeric species, A and B, which are assumed to interact via short-ranged or van der Waals, but not necessarily pairwise-additive, forces.
Such systems are approached in two ways:
(i) via a ``direct,'' ``monomeric,'' or ``physical'' picture, in which the thermodynamics are given solely in terms of the temperature, $T$, and the {\em two} monomeric number densities $\rha$ and $\rhb$; and 
(ii) in terms of an ``associative,'' ``pairing,'' or ``chemical'' picture where the thermodynamics are specified using {\em three} densities, namely, $\rha$ and $\rhb$ for the ``free species,'' and $\rhc$ for the associated pairs, ``complexes,'' or ``two-particle clusters,'' C, {\em along} with a mass action law relating the equilibrium densities corresponding to the reaction \abc.
Most of the principles developed will apply also to ionic fluids; 
the analysis simplifies in the one-component case where A $\equiv$ B and C = A$_2$ \cite{fzlong}.

We show how to match the chemical and physical descriptions explicitly and {\em exactly}.
More concretely, given a ``physical'' Helmholtz free energy density 
$\fbar(T;\rha,\rhb) \equiv - A(T;\rha,\rhb)/k_B T V  $ for a monomeric system of volume $V$, 
we demonstrate how to construct a chemical three-component free energy density, $\fbar_+(T;\rha,\rhb,\rhc)$, which --- when minimized with respect to the adjustable pair density, $\rhc$, for a given association constant, $\kt$ --- precisely reproduces the thermodynamics implied by $\fbar$.
To express this fundamental matching criterion analytically, let 
$\rhc^{\eq} \equiv \rhc(T;\rha,\rhb)$
denote the equilibrium pair density 
determined via the appropriate mass-action law [i.e., by matching chemical potentials or minimizing $A_+(T; \rha,\rhb,\rhc)]$:
then the desired relation is
\begin{equation}
\label{basicmatchabc}
\fbar_+(T;\rha,\rhb,\rhceq) = \fbar(T; \rha + \rhceq, \rhb + \rhceq).
\end{equation}
Although the concept of this precise equivalence is not new (see, e.g., \cite{hill,falkeb,ebgr,kraeft,olaussen}) we are unaware of any similar explicit and exact results for chemical pictures employing only pairs. 

One might feel that if the physical thermodynamics are already known, seeking an equivalent pairing description is of purely academic interest and of little practical use (beyond surmounting the high-temperature difficulties discussed above).
Our primary hope, however, is that the analytical prescriptions discovered may be used as tools and guides for incorporating information that is known exactly or reliably (for example, in terms of some well understood reference system) into {\em approximate} associative thermodynamic descriptions.
Indeed, we are applying the methods to studying the restricted primitive model of electrolytes and its extensions \cite{zfrpm}.

\cplsec{Physical and chemical thermodynamics}

The ``physical'' thermodynamics are given solely in terms of the monomer densities $\rha$ and $\rhb$ which, for clarity, we call $u$ and $v$, respectively, 
so that the overall monomer density is just $\rho = u + v$.
The free energy density  in $d$ spatial dimensions can then be written
\begin{equation}
\label{eqftrue}
\fbar_{\ph}(T;\uu,\vv) = \uu [1 - \ln{(\Lambda_\rma^d \uu)}] 
  + \vv [1 - \ln{(\Lambda_\rmb^d \vv)}] 
  - {\calf}(T;\uu,\vv),
\end{equation}
where we suppose that the excess free-energy density, $F$, embodies the {\em ``true} thermodynamics'' which the chemical picture must match.
As usual, $\Lambda_a(T)$ and $\Lambda_b(T)$ denote the standard thermal de Broglie wavelengths.
For the most part, the temperature dependences will not be displayed since the basic manipulations will be performed at fixed $T$.

In the ``chemical'' or associative picture, the overall monomer density is now
\hbox{$\rho = u+v+2w$,} where $w \equiv \rhc$ is the pair or cluster density.
Correspondingly, we introduce the {\em augmented} or {\em chemical} (reduced) Helmholtz free-energy density via
\begin{eqnarray}
\label{eqfplus}
\fbar_+(T;\uu,\vv,\ww) & = & \uu [1 - \ln{(\Lambda_{\rma}^{d} \uu)}]
  + \vv [1 - \ln{(\Lambda_{\rmb}^{d} \vv)}] 
  + \ww \left\{1 - \ln{[\Lambda_{\rmc}^{2d} \ww /K(T)]}\right\}
	 \nonumber \\
&& \mbox{} + \fbar_+^{\exrm}(T;\uu, \vv, \ww),
\end{eqnarray}
where, to ensure consistency with the classical momentum integrals in the direct partition function, we suppose $\Lambda_c = \sqrt{\Lambda_a \Lambda_b}$ \cite{lf}.
Notice that the association constant, $\kt$, appears here in the role of the internal partition function of an $\text{AB} \equiv \text{C}$ pair \cite{fl,lf,davidson};
for the present purposes, however, $\kt$ may be supposed {\em arbitrary}.

Now, the excess chemical free-energy density may be decomposed according to
\begin{equation}
\label{eqfexdef}
 \fbar_+^{\exrm}(\uu, \vv, \ww) = 
  - {\calf}(\uu,\vv) - \uu \vv {\cale}(\uu,\vv) 
  - \ww {\cald}(\uu,\vv,\ww),
\end{equation}
where the first term, $-F(u,v)$, represents just the physical-picture excess free energy which must be recovered in the absence of association, i.e., when 
$w \equiv \rhc \app 0$.
Then, the ``augmented monomer-interaction function,'' $E(u,v)$, represents the modifications to the A-B interactions inherent in the association process \cite{hill}.
Finally, the ``cluster-interaction function,'' $D(u,v,w)$, embodies the cluster-monomer, A-C and B-C,  and cluster-cluster, C-C, interactions that must be introduced. 

The condition of chemical equilibrium under the reaction \abc $\mbox{}$ follows in standard fashion [by equating the chemical potentials or by minimizing $-\fbar_+(T;u,v,w)$]
leading to the mass-action law,
\begin{equation}
\label{eqmassaction}
w = K u v \gamma_\rma \gamma_\rmb/\gamma_\rmc = K u v e^{M(u,v,w)}, 
\end{equation}
where the activity coefficients, $\gamma_\sigma$, and the excess chemical potential difference, $M(u,v,w)$, are given by
\begin{equation}
\label{eqmuvw}
\ln{(\gamma_\rma \gamma_\rmb / \gamma_\rmc)}
  \equiv M(u,v,w)
  = \left ( \frac{\dee}{\dee \ww} - \frac{\dee}{\dee \uu} 
   - \frac{\dee}{\dee \vv}   \right ) \fbar_+^{\exrm}(\uu,\vv,\ww).
\end{equation}
Of course, (\ref{eqmassaction}) represents a highly nonlinear equation which is normally quite intractable.

\cplsec{Matching via virial expansions}

The free energies for both the physical and chemical pictures may be expanded in powers of monomer and pair densities which permits the establishment of order-by-order matching conditions.
(Compare with \cite{kraeft}.)
One may iteratively solve the mass action relation (\ref{eqmassaction}) to obtain the solution $w = \weq(u,v)$ formally to any desired order.
An expansion of the fundamental matching relation (\ref{basicmatchabc}) solely in terms of $u$ and $v$ may then be obtained by the substitution $w \Rightarrow \weq$.
The $n$th order matching conditions result from equating coefficients of $u^q v^{n-q}$ (for all permissible $q$) in the expansion (\ref{basicmatchabc}).

Letting $F_{lm}, \, E_{lm}$, and $D_{lmk}$ represent the coefficients of $u^lv^m$ in $F$ and $uvE$, and of $u^lv^mw^k$ in $wD$, respectively [see (\ref{eqfexdef})], 
we find, for the two lowest orders,
\begin{eqnarray}
\label{match11}
\; \; \; n=2: \mbox{  \hspace{1.65in} } 
E_{11} & = & K, 
\mbox{ \hspace{2in} }\\
\label{match21}
\; \; \; n=3: \mbox{ \hspace{1in} }
\eeto + K \ddozo & = & \boxhalf K^2 + K ( 2 \fftz + \ffoo ) ,
\mbox{ \hspace{1in} }
\\
\label{match12}
\eeot + K \ddzoo & = & \boxhalf K^2 + K ( 2 \ffzt + \ffoo ) .
\end{eqnarray}
The higher-order relations for both the two- and single-component (A $\equiv$ B) cases have a completely analogous structure and have been obtained explicitly up to $n=5$ \cite{fzlong}.

These relations reveal physically significant features of the chemical picture.
The lowest order relation (\ref{match11}) --- which is implicit, if not more-or-less explicit in much previous work, e.g. \cite{hill,falkeb,lf,kraeft,davidson,stell96} --- admits no adjustable parameters, assuming $K$ is specified.
Since $K$ must be positive, it implies that the {\em chemical} A-B second virial coefficient (the coefficient of $uv$ in $p/k_BT$) 
$B_{110} = F_{11} + E_{11}$, {\em increases} relative to its ``bare'' value, $B_{11} = F_{11}$, in the {\em physical} picture.
In other words, the original A-B interactions must become {\em more repulsive} in the chemical picture.

In next order the relations reveal the freedom implicit in the formulation:
thus in (\ref{match21}), only the {\em combination} ($E_{21} + K D_{101}$) is constrained to match the physically specified, lower-order terms on the right-hand side.
Consequently, the A-C monomer-cluster interaction, which is ``encoded'' in the second virial coefficient $D_{101}$, can be chosen {\em arbitrarily};
likewise, the B-C interaction.
In $n$th order on finds \cite{fzlong}, similarly, that one may choose {\em all} the cluster interaction coefficients $D_{lmk}$ (and hence the C-C interactions, the A-B-C three-body coupling, etc.) {\em arbitrarily} by fixing the corresponding modifications of the bare monomer interactions as embodied in the $n-1$ coefficients
$E_{lm}$ with $l,m \geq 1, \, l+m=n$ \cite{fzlong}.

\cplsec{Exact closed-form chemical formulation}

The latitude discovered in the chemical formulations may, it transpires, be exploited to construct exact, closed-form chemical representations for arbitrary physical thermodynamics.
In particular, the solution of the intractable mass-action relation (\ref{eqmassaction}) may be side-stepped at a comparatively small cost in flexibility \cite{fzlong}.
To this end, let us focus on the excess chemical potential difference, 
$M(u,v,w)$, given by (\ref{eqmuvw}).
If the mass-action equation (\ref{eqmassaction}) had been solved for 
$w = w_{\eq}(u,v)$ --- given the interaction functions $F, \, E$, and $D$ in (\ref{eqfexdef}) --- one could explicitly find $M$ as a function {\em only} of $u$ and $v$.
This suggests that we may {\em choose}, at our discretion, a ``solved'' chemical potential difference function $M_0(u,v)$ [which then represents the {\em exact} equilbrium value of $M(u,v,w)$] and impose the constraint
\begin{equation}
\label{mm0fed}
(\dee_u + \dee_v) \lbrak F(u,v) + uv E(u,v) + w D(u,v,w) 
	\rbrak_{\eq}
		- \dee_w \lbrak w D(u,v,w) \rbrak_{\eq}
= M_0(u,v).
\end{equation}
Here we adopt $\dee_x \equiv \dee/\dee x$ and use the subscript ``eq'' to indicate that one must set $w=\weq(u,v)$ after differentiation.
But $\weq$ may now be obtained trivially: 
by comparing (\ref{mm0fed}) to (\ref{eqmassaction}) and (\ref{eqmuvw}), one sees that the mass-action law has, indeed, been solved in closed form giving just
\begin{equation}
\label{weqmm0}
\weq(u,v) \equiv K u v \exp{[M_0(u,v)]}.
\end{equation}

Now the implications of the constraint (\ref{mm0fed}) are not so obvious;
however, it is linear in $M_0, \, F, \, E,$ and $D$ and, consequently, one finds \cite{fzlong} that much freedom remains explicitly available.
In particular, it proves convenient to specify the cluster-interaction function, $D$, by choosing the ``component functions,'' $D_1$ and $D_2$, in the form 
\begin{equation}
\label{ddgen}
D(u,v,w) = D_0(u,v) D_1(u,v,w) + D_2(u,v,w).
\end{equation}
Physically reasonable and {\em effective} choices for $D_1$ and $D_2$ are discussed below.
The remaining ``coefficient'' $D_0(u,v)$, and the augmented monomer interaction function, $E(u,v)$, can then be expressed simply in terms of $F, \, M_0, \, D_1$, and $D_2$: 
see the Appendix.

Once $D_0(u,v)$ and $E(u,v)$ are calculated via the straightforward prescription in the Appendix, the exact chemical description is complete:
minimizing the free energy (\ref{eqfplus}) with respect to $w$ (i.e., $\rhc$) should now precisely reproduce the original thermodynamics embodied in $F(u,v)$, the physical free energy.

\cplsec{Faithfulness of the exact representations}

Although the closed-form solutions for the chemical free energies constructed in the previous section are formally exact, one can ask:
``Do they actually work?''
To answer this question,
consider, for simplicity, a one-component system (A $\equiv$ B) undergoing the reaction \xxy $\mbox{}$ (although similar considerations apply to the \abc $\mbox{}$ system).
Given a chemical free energy, $\fbar_+(u,w)$, the task of finding equilibrium solutions $u_\eq(\rho)$ and $w_\eq(\rho)$ for a fixed overall monomer density $\rho = u+2w$, can be phrased as seeking the minimum chemical free energy at fixed $\rho$.
Now the exact chemical formulation in Sec.\ 4 guarantees that the mass-action law (\ref{eqmassaction}) is satisfied (with $v \equiv u$);
but that ensures only that $\fbar_+(u,w)$ has an {\em extremum} on the required locus 
$w = K u^2 e^{M_0(u)}$ [see (\ref{weqmm0})].
It is {\em not} guaranteed, however, that this extremum is a free-energy {\em minimum}, let alone the global minimum, as essential for a sensible chemical description.

Thus one may find that not every representation is ``faithful'' --- i.e., exactly reproduces the original, physical thermodynamics --- for {\em all} temperatures and densities.
One may expect that the desired physical solution will be the global free-energy minimum (for arbitrary $D_1$ and $D_2$, say) when the densities $\rho, \, w, \, u$ (and $v$) are sufficiently small;
but failures can, indeed, arise at higher densities for inadequate choices of $D$.

To illustrate this, consider a one-dimensional fluid of hard rods of length $b$ for which the physical ``excluded-volume'' excess free energy is
\begin{equation}
\label{ffrod}
F^{\mathrm{EV}}(u) =-u \ln{(1-bu)}.
\end{equation}
(See, e.g., \cite{lf}.)
To obtain an exact chemical picture with the choice of association constant, say, 
$K=b$, one may try first the ``minimal representation'' generated by the simplest choice, namely,
$M_0=0, \, D_1 = 1, \text{ and } D_2=0$ in (\ref{weqmm0}) and (\ref{ddgen}), with $D_0(u)$ then determined by the one-component versions of (A1)-(A7) \cite{fzlong}.
Numerical and analytical investigations reveal, however, that this ``minimal'' chemical picture remains faithful {\em only} up to 
$\rhostar \equiv \rho b = \rho/\rhomax \simeq 0.26$ \cite{fzlong}:
for larger densities the global chemical free-energy minimum becomes totally unphysical!

Nevertheless, we have found that the situation can be easily remedied by using physical intuition to select a cluster interaction function $D_1$ that reflects the hard-core aspects of the A-C and C-C interactions that one would reasonably expect.
Thus, in parallel to (\ref{ffrod}), the ``excluded-volume'' specification
\begin{equation}
\label{dd1evsub}
{\mathrm HC:} \;\;\;\;\;\;\;
w D_1 = -(u+w)\ln{(1 - b_u u - b_w w)} + u \ln{(1-b_u u)},
\end{equation}
along with $D_2=M_0 = 0$ (as previously) proves very effective.
Explicitly, the natural choice
$b_u = b, \, b_w = 2 b$ is found, by numerical examination, to yield faithful representations up to 
$\rhostar > 0.98$ for assignments of association constants, $K$, ranging from 
$10^{-4}b$ up to $10^4b$.

Furthermore, fully comparable results are obtained for ($d=1$)-dimensional hard-rod A+B mixtures and, in $d=3$ dimensions, for the corresponding Carnahan-Starling \cite{cs,mansoori} hard-sphere ``physical'' thermodynamics (regarded as exact) \cite{fzlong}.
To test and illustrate the theory further we consider systems with attractive interactions.
\pagebreak

\cplsec{Application to a van der Waals fluid}

The van der Waals (vdW) equation gives a surprisingly good semiquantitative description of gas-liquid criticality in simple one-component real fluids which, furthermore, is {\em exact} in an appropriate infinite-range limit \cite{vdwexact}.
The corresponding excess free energy is
\begin{equation}
\label{ffvdw}
F(T,\rho) = -\rho \ln{(1-b\rho)} - b (\eps/k_BT)\rho^2,
\end{equation}
with $b,\,\eps > 0$ so that
convenient reduced units are 
$\rhostar = b \rho$ and $T^* = k_BT/\eps$.
The Boyle temperature, $T_B$, and critical parameters are then
$T^*_B = 1$, $T^*_c = \frac{8}{27}$, 
$\rho^*_c = \frac{1}{3}$.
See Fig.\ 1 for the coexistence curve.

The assignment of an appropriate association constant now calls for some study \cite{fzlong}.
The choice 
$K^{\mathrm vdW}(T) = b \exp{(1/\tstar)}$ matches the temperature dependence of the vdW second virial coefficient at high $T$ --- compare with (\ref{match11}) and Refs. \cite{falkeb,ebgr,dmzfl,fl,lf,davidson} --- 
and also satisfies convexity criteria needed (in approximate representations) to avoid the Second-Law violations referred to above \cite{dmzfl,fzlong}.
It has been adopted for the tests reported here.

Chemical representations of the vdW thermodynamics (\ref{ffvdw}) can be assessed in terms of faithfulness {\em boundaries}, or loci: 
for every temperature, $T$, there will be a density, say $\rho_0(T)$, up to which a given representation is faithful.
As reported for the hard-core models, the minimal formulation ($M_0=D_2=0,\,D_1=1$) fares poorly: see plot (a) in Fig.\ 1.
The results are strongly temperature-dependent below $\tstar \simeq 0.4$ and the critical point lies well {\em outside} the faithful domain!
If one uses the {\em temperature-independent} ``excluded-volume'' cluster-interaction function $D_1$ that that accounts for A-C and C-C repulsions and proved so successful for hard rods --- namely, (\ref{dd1evsub}) with 
$b_u = \frac{1}{2}b_w = b$  --- one certainly improves the behavior at high $T^*$: see Fig.\ 1, plot (b);
but below $\tstar_B$ little is gained.

To obtain better results for low $T^*$, it is reasonable to introduce attractive (i.e., negative) terms into $D$.
Embodying these for convenience only in $D_2(u,v,w)$, we found that the natural form in which $\kt$ regulates the strength of the attractions, namely, 
\begin{equation}
\label{dd2attk}
\;\;\; {\mathrm HCK:} \;\;\;\;\;
w D_2(u,w) = - (d_{21}uw + d_{22}w^2) K(T).
\end{equation}
proves successful.
As evident from plot (c) in Fig.\ 1, imposing only A-C or monomer-cluster interactions, via the choice
$d_{21}=2,\, d_{22}=0$, betters the pure HC representation significantly.
Increasing the A-C coefficient to $d_{21}=3$ yields faithfulness for the full set of tested temperatures in the range $0.05 \leq T^* < \infty$ and densities $0.01 \leq \rho^* \leq 0.99$.
The simple choice 
$d_{21}=d_{22}=1$ with both A-C {\em and} C-C attractions also proves faithful for the entire test set.
This last formulation is probably a reasonable starting point for the chemical description of any model with a choice of association constant that does justice to the monomer attractions at low $T$ \cite{fzlong}.

\cplsec{Concluding remarks}

The exact chemical representations presented above permit --- for the first time --- {\em precise} thermodynamic characterization of the interactions which must arise in an associative, \abc $\mbox{}$ description of a system of A and B monomers.
While monomer-monomer interactions are modified repulsively (since ``paired'' configurations are not permitted ) successful, i.e., faithful, chemical prescriptions explicitly incorporate both repulsive and attractive monomer-pair and pair-pair interactions.
The use of just one representative or dominant type of cluster (i.e., pairs) clearly affects the nature of the required interactions:
the pairs must, in effect, stand in for clusters of all sizes and, at higher densities, may therefore be viewed as ``renormalized,'' ``dressed,'' or ``solvated'' as discussed in \cite{fzlong}.
Except at low densities, therefore, pairs cannot act simply as literal two-body objects if the true ``physical'' or monomeric thermodynamics are to be accurately described.

We hope that the exact chemical formulations will 
prove useful in developing improved associative {\em approximations} for systems with attractive forces.
 The explicit chemical representations found for hard-core reference systems should be of direct use in combination with existing associative approximations for attractions. 
Enforcing the matching relations of Sec.\ 3, moreover, might further refine such theories.
On a slightly different tack --- starting from scratch and drawing solely on the ``thermodynamic'' conception of pairs presented here --- a worthwhile minimal approximation scheme might combine (i) an exact chemical formulation for a reference thermodynamics with (ii) known, low-order virial coefficients as the only further input.
But regardless of the success of particular formulations, knowledge of the exact chemical expressions for simple attractive systems --- which reproduce the known thermodynamics of, say, one-dimensional models or a van der Waals fluid --- may usefully inform the construction of approximate chemical free energies for less tractable systems.
Applications to electrolyte models are underway \cite{zfrpm}.

\vspace{.25in}
\noindent \textbf{Acknowledgments}

The authors thank Professors Harold Friedman and George Stell for their interest in this work, and Dr. Stefan Bekiranov for ongoing discussions and comments on a draft manuscript.
The support of the National Science Foundation (through Grant CHE 96-14495) has been appreciated.

\appendix
\section{Explicit Chemical Representations}
The following expressions \cite{fzlong} complete the chemical representation embodied in (\ref{weqmm0}) and (\ref{ddgen}):
\begin{eqnarray}
D_0(u,v) & = & \left. \lbrak L(u,v) - \dee_w (w D_2) \rbrak_\eq \right/
	\lbrak \dee_w (w D_1) \rbrak_\eq , \\
L[u,v|M_0,{\calf}] & = & 
  \left. \lbrak  J(u,v) + (u+v) \ktwid(u,v) \rbrak \right/ Q(u,v),\\
Q(u,v) & = & 1 + K u v e^{M_0} 
       \lpar \frac{\dee M_0}{\dee u} + \frac{\dee M_0}{\dee v} \rpar 
       + K (u+v) e^{M_0}, \\
J(u,v) & = & \frac{\dee F}{\dee u} + \frac{\dee F}{\dee v}
  + u v \lpar  \frac{\dee \ktwid}{\dee u} + \frac{\dee \ktwid}{\dee v} \rpar
  - M_0(u,v), \\
\ktwid(u,v)  & =  & K [1-M_0(u,v)]e^{M_0} + S \! \lpar u,v;K e^{M_0} \rpar
  + \delf \! \lpar u,v;K u v e^{M_0} \rpar, \\
S(u,v;x) & = & u^{-1} (1 + ux) \ln{(1+ux)} 
	+ v^{-1} (1+vx) \ln{(1+vx)} - 2 x, \\
\Delta F(u,v;x) & = & [ F(u+uvx,v+uvx) - F(u,v) ] / uv. 
\end{eqnarray}
The analogous results for the 2A $\rlharp$ C reaction are given in Ref. \cite{fzlong}.

\begin{figure}[b]
\centerline{ \epsfxsize = 6in
\epsffile{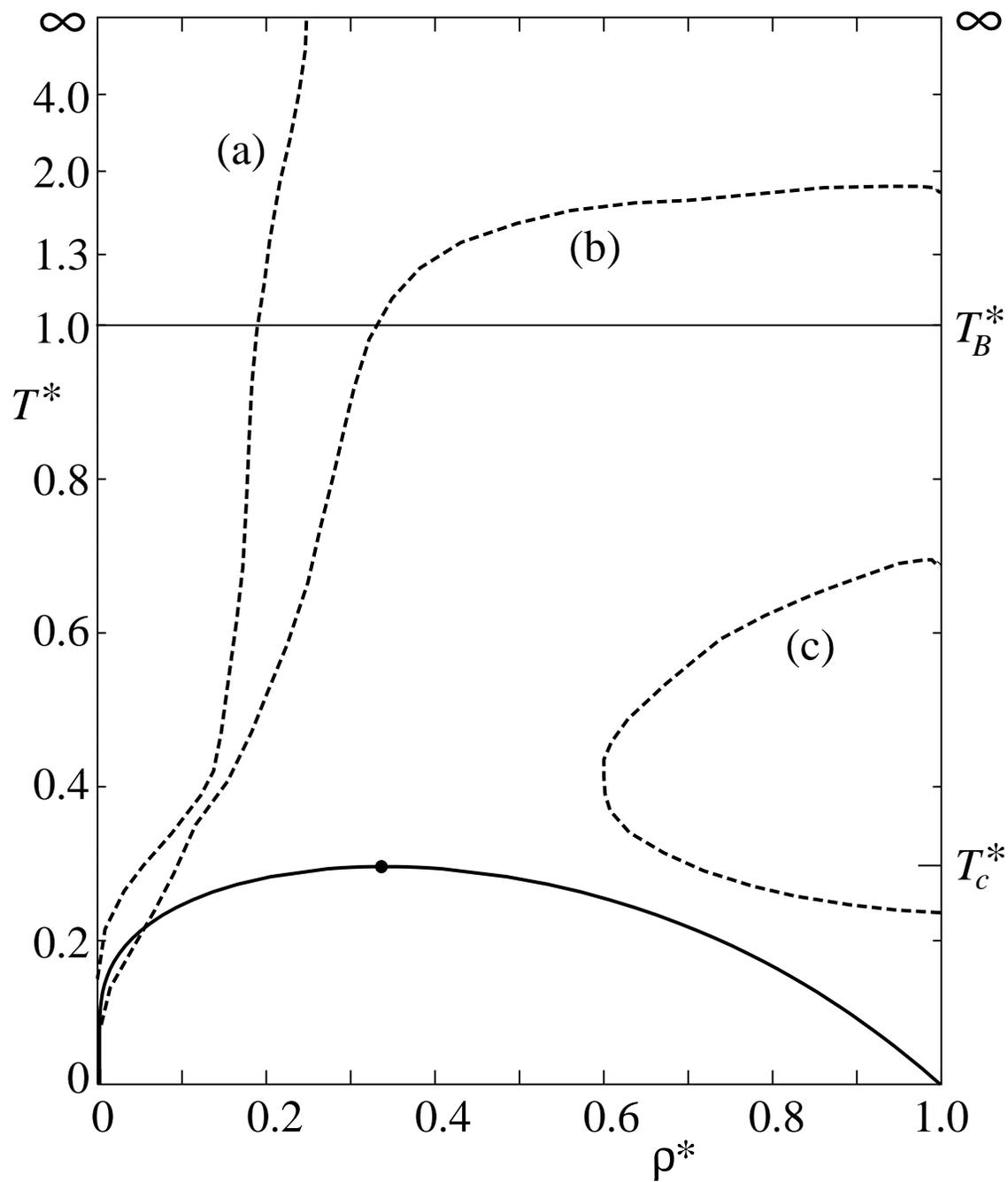}}
\caption{Density-temperature plane for a van der Waals fluid with loci indicating the faithfulness boundaries for various chemical representations: (a) minimal, (b) ``HC,'' and (c) ``HCK'';
see text for details.
Note the nonlinear temperature scale above the Boyle temperature at 
$\tstar_B = 1$.}
\end{figure}

\end{document}